\pdfoutput=1
\documentclass[aps, reprint, superscriptaddress, nofootinbib, showpacs, floatfix]{revtex4-1}
\usepackage{amsmath, latexsym, amssymb, hyperref, graphicx, slashed, color, multirow}
\usepackage[T1]{fontenc}
\definecolor{nicered}{rgb}{.7,.1,.1}
\definecolor{nicegreen}{rgb}{.1,.5,.1}
\definecolor{darkblue}{rgb}{0,0,.5}
\hypersetup{colorlinks, citecolor=nicegreen,linkcolor=nicered, urlcolor=darkblue}

\usepackage{cancel}
\usepackage{amsmath}
\usepackage{amsfonts}
\usepackage{amssymb}
\usepackage{graphicx, rotating}
\usepackage{epstopdf}
\usepackage{epsfig}
\usepackage{latexsym}
\usepackage{graphicx}
\usepackage{color}
\usepackage{amsmath,amssymb}
\usepackage{slashed}
\usepackage{hyperref}

\newcommand{\eq}[1]{Eq.~(\ref{#1})}
\newcommand{\lag}{\mathcal{L}}
\newcommand{\op}{\mathcal{O}}

\newcommand{\be}{\begin{equation}}
\newcommand{\ee}{\end{equation}}
\newcommand{\bea}{\begin{eqnarray}}
\newcommand{\eea}{\end{eqnarray}}
\newcommand{\bc}{\begin{center}}
\newcommand{\ec}{\end{center}}

\def\lra#1{\overset{\text{\scriptsize$\leftrightarrow$}}{#1}}

\begin{document}
\addtolength{\belowdisplayskip}{-.5ex}
\addtolength{\abovedisplayskip}{-.5ex}
\vspace*{-2cm}
\begin{flushright}
CERN-TH-2017\\
Saclay-t18/090
\end{flushright}

\title{$ZZ$ and $Z\gamma$ still haven't found what they are looking for}

\author{Brando Bellazzini}
\email{brando.bellazzini@ipht.fr}
\affiliation{Institut de Physique Th\'eorique, Universit\'e Paris Saclay, CEA, CNRS, F-91191 Gif-sur-Yvette, France}
\author{Francesco Riva}
\email{francesco.riva@unige.ch}
\affiliation{D\'epartment de Physique Th\'eorique, Universit\'e de Gen\`eve,
24 quai Ernest-Ansermet, 1211 Gen\`eve 4, Switzerland}


\begin{abstract}
\noindent
Neutral diboson processes are precise probes of the Standard Model (SM) of particle physics, which entail high sensitivity to new physics effects.  
 We identify in terms of dimension-8 effective operators the leading departures from the SM that survive in  neutral diboson processes  at high-energy, 
and which interfere with the unsuppressed SM helicity contributions. 
We describe symmetries and selections rules that single out those operators, both for weakly and strongly coupled physics beyond the SM. Finally, we show that unitarity and causality enforce, via dispersion relations,  positivity constraints on the coefficients of these effective operators, reducing the parameter space which is theoretically allowed.

\end{abstract}


\maketitle

\section{Motivation}

Precise measurements at the LHC and other colliders are important searching tools for physics beyond the Standard Model (BSM), while also providing a way to test our knowledge of SM processes. 
The well-established context in which these measurements are studied, is that of effective field theories (EFTs), in which departures from the SM are organised as an expansion in inverse powers of a scale $\Lambda$, associated with the physical mass of putative new heavy resonances which are not directly accessible, 
\begin{equation}\label{leff}
\lag^{eff}=\lag^{SM}+\sum_i c^{(6)}_i \frac{\op^{(6)}_i}{\Lambda^2}+\sum_i c^{(8)}_i \frac{\op^{(8)}_i}{\Lambda^4}+\cdots 
\end{equation}
with $\op_i^{(d)}$ operators  of increasing dimensionality $d$, and dimensionless coefficients $c^{(d)}_i$. 

In \emph{most} cases, the leading effects stem from $d=6$ operators, and the series can be truncated there. Natural exceptions come in two kinds. First, there can be BSM scenarios that imply selection rules (e.g. because of symmetries in the underlying theory), such that all $d=6$ operators vanish: in this case the leading effects might arise at $d=8$ \cite{Liu:2016idz,Bellazzini:2017bkb,Contino:2016jqw}, or even higher~\cite{Cheung:2016drk}. Secondly, given that the number of $d=6$ operators is finite, there can be processes which do not receive any contributions at order $d=6$, but are affected only by operators of higher dimensionality.
Neutral diboson processes $pp\to ZV$ ($V=Z,\gamma$)~\cite{CMS:2017ruh,ATLAS:2017eyk}, are  one of the most interesting examples of the latter: they are in fact traditionally interpreted as measurements of neutral triple gauge couplings (nTGC) \cite{Gounaris:1999kf}, which correspond to $d=8$ operators in the EFT language~\cite{Degrande:2013kka}.\footnote{There are also $d=6$ operators modifying these processes (for instance operators of the form $(\bar{\psi} \gamma^\mu \psi)(i H^\dagger \lra D_\mu   H)$), but their contribution is suppressed at high-energy $E$ by powers of $m_Z/E$; in addition, these operators are well constrained by resonant single-$Z$-boson processes (e.g. at LEP) and their impact in $ZZ,Z\gamma$ is negligible.} There are other observables that are first modified  at $d=8$, but are more difficult to identify~\cite{Brivio:2013pma,Gupta:2014rxa}.
Despite the high dimensionality of the operator, the coefficient $c_i^{(8)}$ can  be sizeable  in theories with a relatively strong underlying coupling, so much as to partly alleviate the $E/\Lambda$ suppression that accompanies the contribution of these operators to physical amplitudes. For this reason we can expect measurable effects even from these higher-dimensional operators.

In this work, we discuss the dimension-8 operators that affect $ZZ$ and $Z\gamma$ processes, and propose an innovative way of studying those which is appealing both from an experimental and from a theoretical point of view.
Indeed, effects that are unsuppressed at high-energy, and possibly interfere with the SM in simple analyses, have the highest chance of being detected: these are the effects that are most interesting from an experimental perspective. 
Moreover, since these effects appear in well-motivated and well-structured BSM scenarios, they are interesting from a more theoretical perspective as well, as they represent entire classes of theories. 

In what follows (section \ref{sec:BSMEFT}) we propose a set of operators with all these properties:
  they modify $ZZ$ and $Z\gamma$ amplitudes  at $O(E^4/\Lambda^4)$ (that is the unsuppressed behaviour expected from dimensional analysis); some of them contribute to amplitudes with diboson $+-/-+$ helicities (which happen to be the dominant configurations in the SM, see section~\ref{sec:SM}); finally, most of these operators can be generated at tree-level in models with spin-2 resonances, or arise in scenarios with non-linear supersymmetry, where the SM fermions are Pseudo-Goldstini \cite{Bellazzini:2017bkb}, as we discuss in section \ref{sec:BSM}.

An interesting and curious aspect of certain effective field theories, is the notion of \emph{positivity} which follows, via dispersion relations due to analyticity (causality), Lorentz-invariance, and locality,   from the simple requirement that the underlying microscopic theory be unitary, see e.g.~\cite{Adams:2006sv,Bellazzini:2016xrt}. These positivity bounds do not hold generically, without further assumptions (see \cite{Bellazzini:2014waa,Low:2009di,Falkowski:2012vh}), for dimension-6 operators: for this reason they have not received much attention in contemporary EFT LHC phenomenological studies.
They do imply, however, strict positivity of certain coefficients of our dimension-8 operators, as we discuss in section~\ref{sec:pos}. This model-independent reasoning will provide an important way to focus experimental searches to a smaller region of parameter space.

\section{Standard Model Anatomy}\label{sec:SM}

A thorough understanding of the SM amplitude is necessary to assess what is (and can be) measured at colliders.
We discuss here the partonic $2\to2$ amplitude, which is the target of the simplest inclusive analyses.
The SM tree-contribution to the $\bar\psi\psi\to ZZ,Z\gamma$ processes, is characterised by a $t$($u$)-channel singularity structure, that projects on states of arbitrary angular momentum~$J$, and is dominated by the transverse-transverse (TT) $+-/-+$ helicity amplitudes. Final states with equal helicity $++/--$ are suppressed by  $~m_Z^2/E^2$ ($E=\sqrt{s}$) at high energies~\cite{Azatov:2016sqh}.\footnote{In the SM, dibosons are dominantly produced by quark-antiquark collisions with  subleading quark-gluon and 1-loop  gluon-induced $gg\to VV$ components~\cite{Alioli:2016xab}; we do not consider here these effects.}

The longitudinal-transverse (LT) configuration is instead always suppressed in the high-$E$ limit by $~m_Z/E$. This can be easily understood by noticing that, in the limit of vanishing Yukawas (which makes sense for the type of processes we are considering), a $Z_2$ symmetry $H\to -H$ characterises the SM Lagrangian, implying that amplitudes with an odd number of scalars (that include the Higgs or the longitudinal components of vectors in the high-energy limit) must be suppressed by a vev $v$ and, by dimensional analysis, lead to the above factor~$~m_Z/E$.

Finally, for $ZZ$ the longitudinal-longitudinal (LL) helicity is very small in the tree-level SM, suppressed by  $~m_Z^2/E^2$. This follows from the $t$($u$)-channel SM structure that characterises  tree-level $ZZ$ production in the SM, where the direct coupling of scalars (equivalent to the longitudinal $Z$ polarisations at high energy) to light quarks is suppressed by their small Yukawas.\footnote{The situation is different at NLO, where a gluon-initiated, top-loop mediated diagram, contributes sizeably to the LL final state.}

The amplitudes that do not vanish at high-energy are,
\begin{eqnarray}\label{ampssm}
 {\cal A}^{\textrm{SM}}_{\bar\psi_+\psi_-\to Z^+Z^-}=2g_{Z\psi_-}^2\tan\frac{\phi}{2}+\cdots\\
 {\cal A}^{\textrm{SM}}_{\bar\psi_+\psi_-\to Z^+\gamma^-}=2g_{Z\psi_-}g_{\gamma\psi}\tan\frac{\phi}{2}+\cdots
\end{eqnarray}
where dots denote terms suppressed by powers of $m_Z/E$, $\phi$ is the angle between the momentum of the incoming $h=+1/2$ helicity fermion (or antifermion) and outgoing $h=+1$ helicity vector (that can be a $Z$ or $\gamma$), and the subscript $\pm$ denotes the sign of the helicity. Here  $g_{Z\psi}=g(T^3_\psi-\sin^2\theta_WQ_\psi)/\cos^2\theta_W$ and $g_{\gamma\psi}=eQ_\psi$. Amplitudes with opposite vector helicity are simply related to those of \eq{ampssm} by
\begin{equation}
 {\cal A}^{\textrm{SM}}_{\bar\psi\psi\to Z^+V^-}(\phi)= {\cal A}^{\textrm{SM}}_{\bar\psi\psi\to Z^-V^+}(\pi+\phi),
\end{equation}
(with $V=Z,\gamma$) and similarly for  $\psi_+\leftrightarrow \psi_-$ interchange, keeping in mind that the $Z$ couplings to fermions are chiral $g_{Z\psi_+}\leftrightarrow g_{Z\psi_-}$.

A final important piece of information, is the fact that the largest SM amplitude is associated with left-handed initial state quarks, due to the known suppression of the right-handed quarks coupling to the $Z$-boson: $\{g_{Zd_L},g_{Zu_L},g_{Zu_R},g_{Zd_R}\}\sim\{-0.32,0.27,-0.10,0.05\}$. 

\begin{figure}[htb]
\begin{center}
\includegraphics[width=0.45\textwidth]{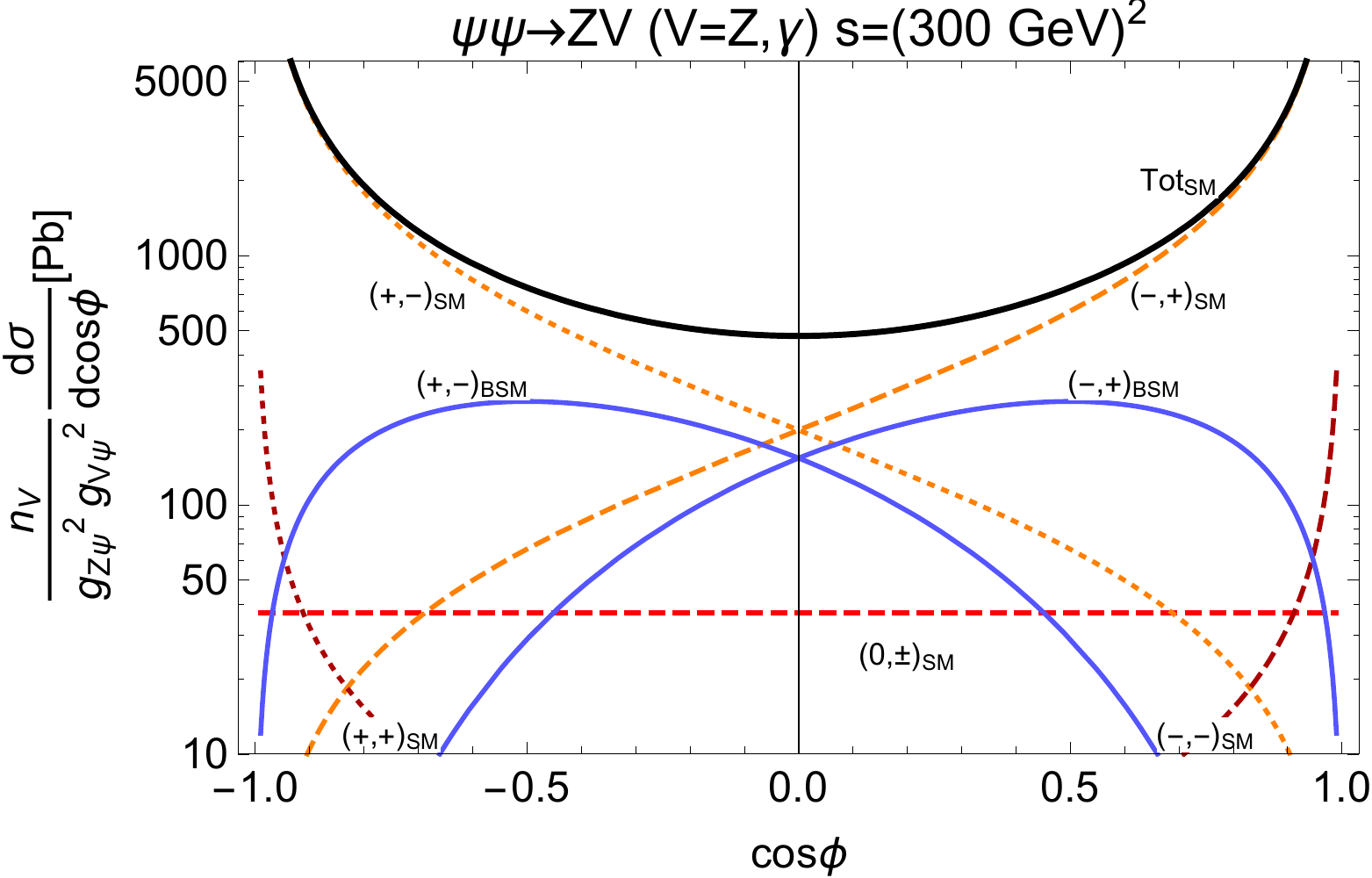}
\end{center}
\caption{\it Differential $\bar\psi\psi\to ZZ,Z\gamma$ cross-section for different helicities, in units of the $\gamma\psi,Z\psi$ couplings, $g_{\gamma\psi}, g_{Z\psi}$ ($n_V=1$ for $Z\gamma$ and $n_V=2$ for $ZZ$). Dashed coloured lines correspond to the SM-only contribution (solid black, the sum over helicities), while solid blue to the BSM-only TT polarisations, with an arbitrarily chosen normalisation, to be shown in the same plot as the~SM.}\label{fig:dsit}
\end{figure}

The content of this section is summarised in Fig.~\ref{fig:dsit}. This can be trivially extended to $\bar\psi\psi\to\gamma\gamma$ processes, which we do not discuss in detail, as it is not traditionally discussed within the nTGC framework.

\section{Effects Beyond the Standard Model}\label{sec:BSMEFT}

The nTGC parametrization of Ref.~\cite{Gounaris:1999kf} , and its EFT counterpart \cite{Degrande:2013kka}\footnote{Notice that  anomalous-couplings always induce an energy growth in some scattering process, that ultimately implies a cut-off at energy $E\sim\Lambda^{cutoff}$, above which the associated theory ceases making sense: for this reason anomalous coupling parametrizations can always be reformulated as EFTs with $\Lambda\leq\Lambda^{cutoff}$.}, is based on the physical hypothesis that the putative underlying new dynamics only  modifies interactions among 3 gauge bosons. Consequently, new physics can  enter $\bar\psi\psi\to ZV$ only through an s-channel diagram or, more concretely, only the $J=1$ amplitude can be modified. Because of angular momentum selection rules, only a Longitudinal+Transverse (LT) diboson final state is then possible for the neutral SM gauge bosons~\cite{Jacob:1959at}. This amplitude is suppressed by $m_Z/E$ also in BSM, and its interference with the SM is even more suppressed, since the SM LT piece is small.

Here we take a step forward and ask ourselves how $\pm\mp$ diboson helicities can be sourced.  The first configuration that allows for this is $J=2$, and is sourced, for initial states involving fermions, by an operator containing $\left(i\bar \psi \gamma^{\{\mu} D^{\nu\}} \psi+\mathrm{h.c.}\right)$.
From this, the leading CP even operators that we can write and lead to a $ZZ,Z\gamma$ final state are
\begin{align}
\op^{(8)}_{\psi B}=& -\frac{1}{4}\left(i \bar \psi \gamma^{\{\rho}D^{\nu\}} \psi+\mathrm{h.c.}\right)B_{\mu\nu}B^{\mu}_{\phantom{\mu}\rho}\label{ob}\\
 \op^{(8)}_{\psi W}=&-\frac{1}{4}\left(i \bar \psi \gamma^{\{\rho}D^{\nu\}} \psi+\mathrm{h.c.}\right) W^a_{\mu\nu}W^{a\,\mu}_{\phantom{\mu}\rho}\label{ow}\\
 \op^{(8)}_{\psi H}=&\frac{1}{2}\left(i\bar \psi \gamma^{\{\mu} D^{\nu\}} \psi+\mathrm{h.c.}\right)D_\mu H^{\dagger} D_\nu H\label{oh}
\end{align}
in the neutral channel, and
\begin{align}
\hat\op^{(8)}_{Q H}=&\frac{1}{2}\left(i\bar Q \sigma^a\gamma^{\{\mu} D^{\nu\}} Q+\mathrm{h.c.}\right)D_\mu H^{\dagger} \sigma^a D_\nu H \label{oh1}\\
\hat\op^{(8)}_{Q BW}=&-\frac{1}{4}\left(i \bar Q \sigma^a\gamma^{\{\rho}D^{\nu\}} Q+\mathrm{h.c.}\right)B_{\mu\nu}W^{a\,\mu}_{\phantom{\mu}\rho}\label{owb}
\end{align}
in the isospin-charged  channel; here $\psi=Q_L,u_R,d_R$ denotes the initial state fermion relevant for LHC. These operators are dimension-8, that is, they are of the same order as effects that are tested at present as anomalous~nTGC~\cite{Degrande:2013kka}.

The interactions in Eqs.~(\ref{ob},\ref{ow},\ref{owb}) modify the TT amplitudes, contributing with different combinations to $Z_TZ_T$ as well as $Z_T\gamma$ and $\gamma\gamma$ final states. For example, from Eqs.~(\ref{ob},\ref{ow}), at high-energy we find for $\psi=u_R$ or $d_R$ 
\begin{align}\label{growingtrans}
 {\cal A}^{\textrm{BSM}}_{\bar\psi_+\psi_-\to Z^+V^-}=\frac{c^{tot}_{V}}{4}\frac{\hat s^2}{\Lambda^4} \sin \phi\,\, (1+\cos\phi)\,,
\end{align}
where $c^{tot}_{Z}=\sin^2\theta_Wc^{(8)}_{\psi B}+\cos^2\theta_Wc^{(8)}_{\psi W}$ and $c^{tot}_{\gamma}=\sin(2\theta_W)(c^{(8)}_{\psi B}-c^{(8)}_{\psi W})$ are the effective combinations that enter in the two processes ($\theta_W$ the weak mixing angle). Here $\hat s$ is the center-of-mass energy, and \eq{growingtrans} exhibit the unsuppressed energy-growth that one expects from dimension-8 effects. The associated differential cross sections are shown in Fig.~\ref{fig:dsit}.  Analogous expressions hold for initial states with $\psi=Q$, and $\gamma\gamma$ final states.

Similarly, Eqs.~(\ref{oh},\ref{oh1}) contribute at high-energy $E\gg m_Z$ to the production of two longitudinally polarized $Z$ bosons, as can be understood by the equivalence theorem and dimensional analysis. From \eq{oh} we find at high-energy,
\begin{equation}
 {\cal A}^{\textrm{BSM}}_{\bar\psi_+ \psi_- \to Z_L Z_L}=\frac{c^{(8)}_{\Psi H}}{8}\frac{\hat s^2}{\Lambda^4} \sin (2\phi)\,.
\end{equation}

An important aspect of these precision SM tests is the interference between the SM process and the BSM effect: a sizeable interference enhances the sensitivity to these BSM effects.
As discussed in the previous section, the majority of SM processes that we observe at the LHC have $\pm\mp$ helicity, and left-handed initial state quarks. This amplitude is modified by the operators of Eqs.~(\ref{ob},\ref{ow},\ref{owb}) with $\psi=Q_L$. Furthermore, as can be observed from Fig.~\ref{fig:dsit}, the SM and BSM same-helicity polar-angle distributions have an important overlap, implying that a sizeable SM-BSM interference can be expected in fixed energy bins.
Simple analysis with standard selection criteria, including a small cut in the very-forward region $|\cos\phi|\approx 1$, shall be already particularly sensitive to  the deformations of Eqs.~(\ref{ob},\ref{ow},\ref{owb}).

Interference of the LL channel is instead suppressed, given the smallness of the LL SM amplitude (see previous section).
So this channel maintains its interest mainly from its very-high-energy behaviour and the BSM connection with modified Higgs dynamics.

On the other hand, it is interesting to notice that  analyses based on the LT configurations (as implicitly assumed  by  nTGC analyses) imply that this majority of SM events plays effectively the r\^ole of background, rather than signal. Instead, the operators of Eqs.~(\ref{ob},\ref{ow},\ref{owb}) source the right helicity amplitudes so that the situation is reversed and all available experimental information is systematically used and tested.

The final states we are interested in are not unique of the $J=2$ angular momentum configuration and are also found for $J\geq 3$. However states with larger angular momentum are necessarily sourced by local operators with additional powers of momenta/derivatives. That is, $J\geq3$ is associated with operators of dimension $>8$ that are, in this situation, negligible.

\section{BSM Perspective}\label{sec:BSM}

In the previous sections we have motivated, from an experimental (bottom-up) point of view, a new class of effects that appears at the same order in the EFT expansion as effects that are currently studied, but are easier to detect (for a comparable new physics scale) given their unsuppressed nature and their interference with the~SM. Now we take a more theoretical (top-bottom) perspective and argue that these operators are also the ones favoured from a BSM point of view: we discuss two BSM scenarios in which these effects could arise with sizeable coefficients.

Here it is perhaps worth pausing a moment and understand whether dimension-8 operators can be relevant at all, given that they appear to be subleading in the energy expansion.
A given BSM that generates sizeable $d=6$ effects as well as $d=8$ ones, will probably  be better searched through those $d=6$ effects. So, we are interested in whether it is possible that $c^{(6)}\ll c^{(8)}$, that is, situations where the coefficients of $d=6$ operators are suppressed while $d=8$ are not. 
This is certainly possible in some finely tuned region of parameter space, but this goes against the perspective that EFTs capture broad BSM scenarios, rather than specific points in parameter space. 
Symmetries and  other selection rules can instead induce natural situations for this hierarchy, as we now discuss with two examples.

A first situation that leads to dimension-8 effects that are larger than dimension-6 ones, is the tree-level exchange of weakly-coupled massive spin-2 resonances, such as graviton Kaluza-Klein excitations in models with extra-dimensions~\cite{Giudice:1998ck,Han:1998sg,Hewett:1998sn}. 
The massive graviton interacts (like its massless version) with the stress-energy tensor~$T_{\mu\nu}$; its Lagrangian is
\begin{equation}
\lag_g=-\frac{m_g^2}{2}h^{\mu\nu}P_{\mu\nu\rho\sigma}h^{\rho\sigma}-\frac{1}{\bar M_p}h^{\mu\nu}T_{\mu\nu}
\end{equation}
with $P_{\mu\nu\rho\sigma}=(\eta_{\mu\rho}\eta_{\nu\sigma}+\eta_{\mu\sigma}\eta_{\nu\rho})/2-\eta_{\mu\nu}\eta_{\rho\sigma}/3+\cdots$, which is equivalent to  the propagator expanded at leading order in momentum over the spin-2 mass $p/m_g$, and $\bar M_p$ the reduced Planck mass in the extra dimension. Integrating out $h$ one finds at leading order in $1/m_g$
\begin{equation}
\lag^{eff}_{g}=\frac{1}{2m_g^2\bar M_p^2}[(T^{\mu\nu} T_{\mu\nu})-\frac{1}{3}(T^\mu_\mu)^2]+\cdots\,.
\end{equation}
The second piece only leads to effects with off-shell fermions and is not relevant for our discussion (in fact, via a field redefinition, it can be written as a dimension-10 effect),  while the first one leads to 
\begin{equation}\label{gravitoncoeffs}
c_{\psi H}= c_{\psi B}= c_{\psi W}=\frac{m_g^2}{\bar M_p^2} \,\quad
\end{equation}
with $\Lambda=m_g$, in addition to a number of other $d=8$ operators involving four fermions, vectors or scalars \cite{Giudice:1998ck}. 

We have illustrated a model that singles out $d=8$ operators, and in particular the ones proposed in this note.
This holds only in the limit where $m_g$ is much lighter than other BSM resonances, and moreover is  weakly coupled, meaning that the relevant coupling at the scale $m_g$ is $\ll 4\pi$.\footnote{Notice that most LHC processes are at present tested with large statistical uncertainty at high-energy, aiming at deviations from the SM which are larger than~$O(1)$: in this situation, a consistent EFT interpretation is possible only if the underlying theory is coupled more strongly than the SM \cite{Contino:2016jqw}. This suggest a sizeable window of parameters where these models are represented by a $d=8$ EFT and are consistently testable.}
 Indeed for strong coupling, graviton loops generate $d=6$ effects \cite{Giudice:2003tu} and, moreover, one expects a richer spectrum of states at the cut-off.

Symmetries provide instead a more robust context to study the relevance of $d=8$ effects, that can also hold in the strongly coupled limit.
Consider for instance a real scalar field~$\phi$, Nambu-Goldstone boson of a $U(1)$ symmetry, spontaneously broken at the physical scale $\Lambda$. The $U(1)$ symmetry translates into  a shift symmetry $\phi\to \phi+\alpha$ ($\alpha$ a global $U(1)$ phase) which implies that the effective Lagrangian respecting this (non-linearly realized) symmetry can only involve powers of $\partial \phi$:
\begin{equation}\label{philag}
\lag^{eff}_\phi=\frac{1}{2}\partial_\mu\phi\partial^\mu\phi+\frac{c_\phi}{\Lambda^4}(\partial_\mu\phi\partial^\mu\phi)^2 +\cdots
\end{equation}
Clearly, in this theory it is natural that the leading interactions are dimension-8, a statement that is independent on the coupling strength in the microscopic theory,  since the symmetry always protects against the generation of $d=6$ terms. This  remains true even if the $U(1)$ symmetry is only approximate, broken explicitly by small parameters, such as a mass term $m^2_\phi \phi^2/2$, with $m_\phi\ll \Lambda$. In this situation, dimension-6 operators (such as $(\partial_\mu|\phi|^2)^2$) will be generated, but we can expect their coefficients to be small and controlled by the small parameter $m_\phi^2/\Lambda^2\ll 1$.

The very same reasoning has been applied to the SM fields in Refs.~\cite{Giudice:2007fh,Liu:2016idz,Bellazzini:2017bkb}. In particular Refs.~\cite{Bellazzini:2017bkb} discusses the analog of the $\phi\to \phi+\alpha$ shift-symmetry in the context of fermions: non-linearly realized extended supersymmetry. 
If the SM fermions are (pseudo)-Goldstini of spontaneously broken~(SB) supersymmetry, then their leading interactions arise indeed at $d=8$ and provide an example where dimension-8 effects can be well motivated from a BSM perspective.

The low-energy physics of Pseudo-Goldstini can be captured by an effective theory for SB space-time symmetries so that the main couplings between Pseudo-Goldstini $\chi$ and matter, arise effectively as a distortion of the effective metric perceived by matter fields in the SB background,
\begin{equation}
 \label{inversmetexp}
 g^{\mu\nu}=\eta^{\mu\nu}+\frac{1}{2F^2}\left(i\bar\chi \gamma^\mu \partial^\nu \chi+ i\bar\chi \gamma^\nu \partial^\mu \chi+\mathrm{h.c.}\right)+\ldots 
 \end{equation}
 where  $F$ is the supersymmetry breaking scale.
In this context, if the SM fermions are pseudo-Goldstini $\psi=\chi$,  then the kinetic term for Higgs $D_\mu H^\dagger D_\nu H g^{\mu\nu}$ leads to \eq{oh}, while the $V=W,B$ kinetic terms $-1/4 V^A_{\mu\nu} V^A_{\rho\sigma} g^{\mu\rho}g^{\nu\sigma}$ lead to the interactions of Eqs.~(\ref{ob},\ref{ow}), with 
\begin{equation}\label{coeffsgoldstino}
c_{\psi H}=c_{\psi B}=c_{\psi W}=\frac{\Lambda^4}{F^2}\,,
\end{equation}
with $\Lambda$ the mass of the other supersymmetric particles.

The complete set of (approximate) symmetries that can lead to $c^{(8)}\gg c^{(6)}$ has been discussed in Refs.~\cite{Liu:2016idz,Bellazzini:2017bkb}: while non-linear SUSY can protect Eqs.~(\ref{oh}-\ref{ow}), there is no  known analog for other operators entering diboson pair-production. This puts the operators Eqs.~(\ref{oh}-\ref{ow}) on a privileged ground, also from a BSM perspective.

\section{Positivity Constraints and Beyond}\label{sec:pos}

Dispersion relations, following from the fundamental principles of causality (analyticity), locality (Froissart bound~\cite{Froissart:1961ux,Martin:1965jj}), and crossing symmetry,  imply relations between low-energies (IR), where LHC experiments are performed, and high-energies (UV), where the SM and the effects Eqs.~(\ref{oh}-\ref{owb}) are completed into a microscopic unitary theory. This implies strict positivity constraints for the coefficient of the $s^2$ term in the Taylor expansion of elastic scattering amplitudes at $t,s\to0$, 
\begin{equation}\label{pos}
\partial^2 {\cal A}/\partial s^2|_{s,t=0}>0\,.
\end{equation}
This has important consequences for EFTs with unknown, but unitary, UV completions \cite{Adams:2006sv,Bellazzini:2016xrt}.
By dimensional analysis it is clear that this is a unique feature of operators  of dimension $d=8$  (or $d\geq 8$, via higher derivatives in \eq{pos}) that is unparalleled in the more familiar framework of $d=6$ operators (see however Ref.~\cite{Low:2009di,Falkowski:2012vh,Bellazzini:2014waa} for a similar relation involving additional inputs from the UV).

Applied to the operators of Eqs.~(\ref{ob}-\ref{oh}), appearing in the Lagrangian as in \eq{leff}, with $\Lambda^4>0$, we find,\footnote{The following positivity bounds are extracted at very small energy, where the Wilson coefficients differ from those at the scale~$\Lambda$ (e.g. Eqs.~(\ref{gravitoncoeffs},\ref{coeffsgoldstino}). Here we assume that the rest of the  theory (i.e. the SM and possible dim-6 operators) is weakly coupled below $\Lambda$, so that (calculable) RGE effects can be neglected, as long as $c^{(8)}$ is not much larger than $(4\pi)^2$, because otherwise  insertions of operators with $d<8$ and at least one dimension-8 operator with such a large coefficient might generate large radiative corrections.}
\begin{equation}\label{positivitycons}
c^{(8)}_{\psi H}>0\,,\quad c^{tot}_{Z}>0\,, \quad c^{tot}_{\gamma\gamma}>0 \,, 
\end{equation}
for $\psi=u_{R}\,, d_{R}$, and where $c^{tot}_{\gamma\gamma}=c^{(8)}_{\psi B}\cos^2\theta_W + c^{(8)}_{\psi W}\sin^2\theta_W $.
That is, fundamental principles reduce the parameter space that can be explored to  half its size and help focussing experimental searches. These (strictly positive) constraints also imply that these operators are necessarily there and it would not make sense to study, for instance, scenarios where BSM starts with leading $d>8$ operators (or $J>3$ angular momentum) \cite{Bellazzini:2016xrt}. 
Scattering instead $\psi=Q=(u_L,d_L)$ states, the operators in (\ref{oh1}-\ref{owb}) contribute to the forward elastic amplitudes as well, the results varying with the isospin $\sigma^3_{ii}=\pm 1$ associated to $u_L$ and $d_L$ respectively,
\begin{align}
{\cal A}_{Q Z_L \rightarrow Q Z_L}(s,t=0) & = \left( c^{(8)}_{QH}\mp \hat{c}^{(8)}_{QH}\right)  \frac{s^2}{\Lambda^4} \,, \\ 
{\cal A}_{Q V \rightarrow Q V}(s,t=0) & =c^{(8)\pm}_{Q V}   \frac{s^2}{\Lambda^4}\,.\label{eqqvqv}
\end{align}
Here $V=Z_T,\gamma$, and we defined $c^{(8)\pm}_{Q Z}=\sin^2\theta_W c^{(8)}_{Q B}+\cos^2\theta_Wc^{(8)}_{Q W} \mp \hat{c}^{(8)}_{Q BW}\sin(2\theta_W)/2$, 
$c^{(8)\pm}_{Q \gamma}=c^{(8)}_{QB}\cos^2\theta_W + c^{(8)}_{QW}\sin^2\theta_W \pm \hat{c}^{(8)}_{QBW}\sin(2\theta_W)/2 $.
Therefore, positivity implies that
\begin{equation} \label{poscondtheta}
c^{(8)}_{QH}\mp \hat{c}^{(8)}_{QH}> 0 \,, \qquad c^{(8)\pm}_{Q V}   >0\,,
\end{equation}
and that $\hat c^{(8)}_{Q BW}$ cannot appear without (equally large) contributions from $c^{(8)\pm}_{Q V}$. 

\begin{figure}[tb]
\begin{center}
\includegraphics[width=0.4\textwidth]{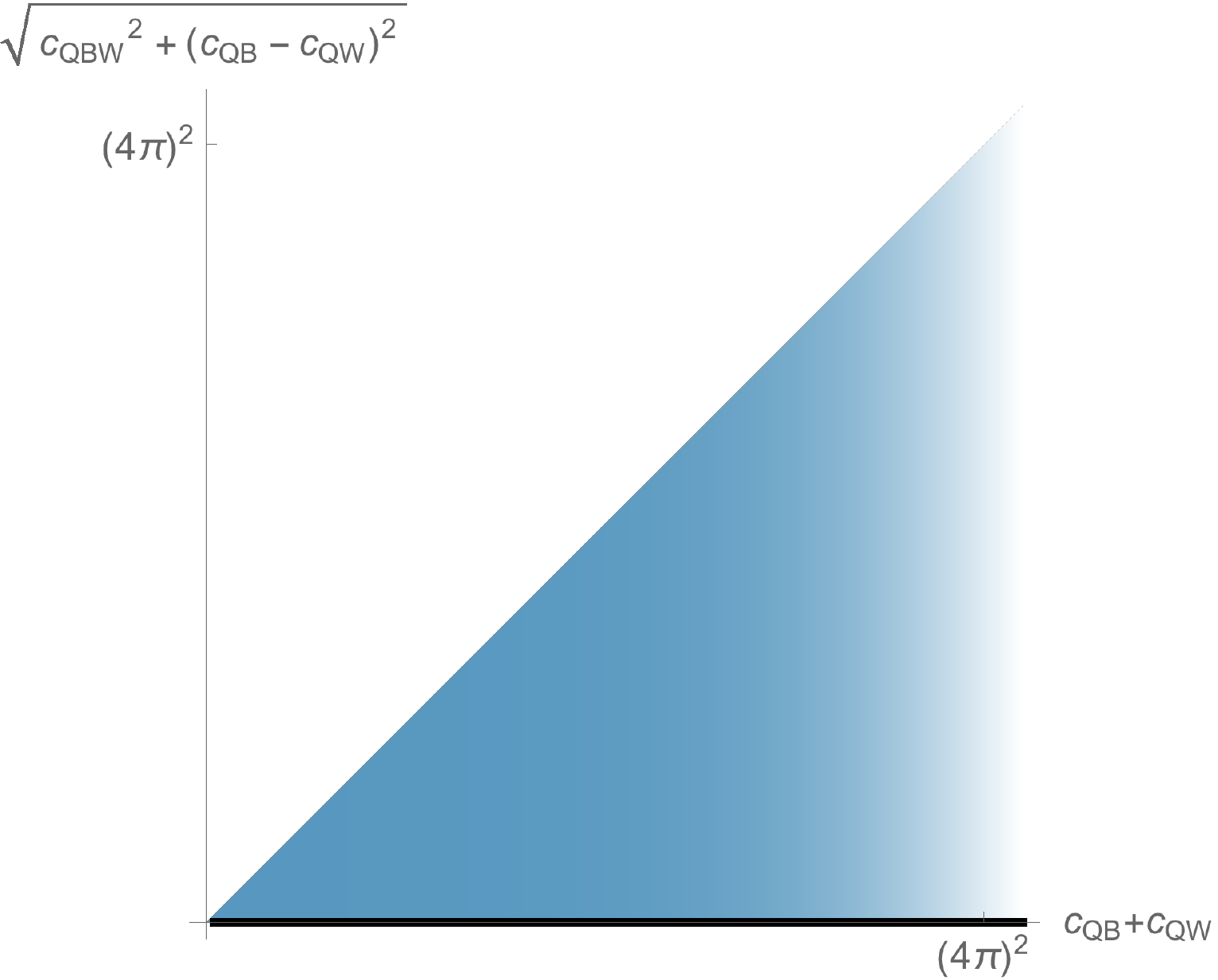}
\end{center}
\caption{\it Parameter space for the operators with $\psi=Q$ (we have neglected  superscripts for clarity). The blue region is allowed by positivity constraints Eqs.~(\ref{posconst},\ref{poscone}).
The black segment correspond to the explicit scenarios of section \ref{sec:BSM}. 
}\label{fig:pos}
\end{figure}

The  statement of \eq{poscondtheta} can be made even stronger if the dynamics that generates these operators 
is insensitive to the low-energy electroweak~(EW) and EWSB physics, that is if the new sector admits a consistent UV completion irrespectively of the Weinberg mixing angle $\theta_W$ which is determined by the external weak gauge couplings.\footnote{Examples of this are the KK graviton or Goldstini of  section~\ref{sec:BSM}, where the transverse gauge bosons are also part o the strongly interacting sector, as described in Ref.~\cite{Liu:2016idz}.}  More explicitly, we are advocating for the gedanken limit $g\rightarrow 0$ and $g^\prime\rightarrow 0$ while allowing the ratio $g^\prime/g=\tan\theta_W$ to be free and independent of the strong sector. This means that the bounds~(\ref{positivitycons}, \ref{poscondtheta}) hold for any $\theta_W$, 
which in turn imply that \begin{equation}\label{posconst}
c^{(8)}_{\psi H}>0\,,\quad c^{(8)}_{\psi W}>0\,,\quad c^{(8)}_{\psi B}>0 \,,\quad c^{(8)}_{QH}>  |\hat{c}^{(8)}_{QH} |\,.
\end{equation}
with $\psi=Q$, $u_R$ and $d_R$; and additionally 
\begin{equation} \label{poscone}
4 c^{(8)}_{QB}  c^{(8)}_{QW} > (\hat{c}^{(8)}_{QBW})^2 \,.
\end{equation}
The inequalities for $\psi=Q$, which derive from \eq{eqqvqv} using $\theta_W$ as a free parameter, define a one-branch cone in the $(c^{(8)}_{QB}, c^{(8)}_{QW},\hat{c}^{(8)}_{QBW})$ space, which is obtained by revolving the triangle shown in  Fig.~\ref{fig:pos} around the axis $c^{(8)}_{QB} + c^{(8)}_{QW}$. %
These generic results are of course compatible with the specific models discussed above, Eqs.~(\ref{gravitoncoeffs},\ref{coeffsgoldstino}), where $\hat{c}^{(8)}_{QBW}$ and $\hat{c}^{(8)}_{QH}$ are not generated, and all other operators are generated with positive coefficients.

In addition to the above-mentioned  \emph{lower} bounds on the coefficients $c^{(8)}$, there are various arguments that suggest \emph{upper} bounds.  These arguments include perturbativity, beyond-positivity \cite{Bellazzini:2017fep}, or simple Naive Dimensional Analysis~\cite{Manohar:1983md,Cohen:1997rt}, and indicate that for coefficients $c\gg (4\pi)^2$, the EFT description that we have proposed (where only $d=8$ operators are retained), may break down.

%

\section{Conclusions and Outlook}

In this work, we have provided a parametrization for neutral diboson processes with initial state fermions, that extends the traditional parametrization based on neutral triple gauge couplings. 
The scope has been to provide a framework for testing the precise SM predictions \cite{Cascioli:2014yka,Kallweit:2018nyv,Denner:2015fca,Kallweit:2017khh,Grazzini:2015nwa}, against the most interesting and well-motivated alternative hypotheses.
Our arguments have been exposed in terms of an Effective Field Theory (where they are naively of the same order as nTGC effects), but are based on the general helicity properties of diboson amplitudes. In particular, we point out $J=2$ effects that are unsuppressed at high energy, both in the LL and TT final states, and highlight effects that do interfere with the dominant SM amplitude which have $\pm\mp$ helicity.  In this sense our proposal goes towards the goal of exploiting at best LHC data, providing a way to present information about the high-energy SM amplitudes in its completeness, and identifying the features that can be tested most precisely.

We have then approached these operators from a more theoretical perspective, and argued that fundamental principles based on unitarity imply generic and model-independent constraints on the operator coefficients. This is \emph{per se} a very interesting result that has no general analog in the more familiar context of $d=6$ operators. As a result, the parameter space of the naive EFT is drastically reduced, thus focussing experimental efforts to the relevant physical scenarios. We find for instance that the operators of Eqs.~(\ref{oh1},\ref{owb}), when taken in isolation, cannot be completed into a unitary theory, despite being allowed from a na\"ive EFT point of view.

Finally, we have motivated the proposed effects from a BSM point of view. We have provided a specific model where the virtual tree-level exchange  a massive spin-2 resonance (KK graviton) generates the operators under scrutiny;  $d=6$ effects are loop-suppressed, so that in the weakly coupled regime this example produces its larger effects in our $d=8$ operators.
Moreover, we have identified a symmetry (non-linearly realized supersymmetry) that singles out  the operators Eqs.~(\ref{ow}-\ref{oh}) and holds also in the strongly coupled regime. This puts the hierarchy $c^{(8)}\gg c^{(6)}$ of our scenario on a firm ground both in the realm of weakly and strongly coupled models.

This opens several interesting prospects for future research. From an experimental point of view, it will be interesting to asses the reach of LHC experiments to this type of physics (in particular to the operators of Eqs.~(\ref{ow}-\ref{oh}) with $\psi=Q_L$), with present and future data, and eventually implement this strategy into present $ZZ$ and $Z\gamma$ studies. Contrary to charged diboson processes, $ZZ,Z\gamma$ give easy access to the center-of-mass energy $\hat s$, facilitating therefore a discussion of the EFT validity, along the lines of Ref.~\cite{Contino:2016jqw}.
Moreover, resonance searches for spin-2 (KK graviton) states are already performed at the LHC in the $ZZ$, $Z\gamma$ or $\gamma\gamma$ channels \cite{Sirunyan:2016cao,ATLAS:2017spa,Aad:2015mna,Khachatryan:2016yec}: it would be nice to compare and combine these complementary modes of exploration (resonant and EFT) into a single unified picture.
Finally it would be interesting to discuss NLO effects, and in particular their contribution to the $00$ amplitude to understand whether this can be exploited to search for specific BSM scenarios, e.g. \cite{Bellazzini:2015cgj,Goncalves:2018pkt}, in the $gg\to ZZ$ amplitude.

From a model-building point of view, the Pseudo-Goldstino models of Ref.~\cite{Bellazzini:2017bkb} can be put on firmer ground, perhaps even their relation to the hierarchy problem, thus providing additional motivation for these searches.
Finally, the reasoning we develop here could be extended to charged diboson final states $W^\pm Z$, $W^+W^-$.
Indeed, it has been shown on completely general grounds, that $d=6$ effects including transverse vectors have suppressed SM-BSM interference in the high-energy limit~\cite{Azatov:2016sqh} for inclusive searches (see however~\cite{Panico:2017frx}): in these conditions the effects discussed here and those of the dimension-6 operator $W^3$ would appear at the same order, a feature that might deserve a dedicated discussion in explicit scenarios.

Our proposal offers an innovative opportunity for experiments transiting towards an EFT parametrization of non-SM effects.

\vspace{1cm}
\noindent 
{\bf Acknowledgements.} 
We acknowledge important conversations with Javi Serra, Francesco Sgarlata, Alex Pomarol, and Andrea Wulzer. We thank Javi Serra and Francesco Sgarlata for the the involvement at the early stage of this project.  B.B. thanks Marco Cirelli, Roberto Contino, and Enrico Trincherini for the kind hospitality at the LPTHE and at the SNS.
FR is supported by the Swiss National Science
Foundation under grant no.  PP00P2-170578.

\def\hhref#1{\href{http://arxiv.org/abs/#1}{#1}} 

\end{document}